\documentclass[preprint,aps,tightenlines,showpacs,nofootinbib]{revtex4}
\usepackage{latexsym}
\usepackage[dvips]{graphicx}
\newcommand{\beq}{\begin{eqnarray}}
\newcommand{\eeq}{\end{eqnarray}}
\newcommand{\eq}{eqnarray}

\newcommand{\ci}{\cite}

\newcommand{\de}{{\delta}}

\newcommand{\La}{{\Lambda}}

\newcommand{\Om}{{\Omega}}
\newcommand{\pa}{{\partial}}
\newcommand{\no}{{\nonumber}}
\newcommand{\f}{\frac}
\newcommand{\ra}{\rightarrow}

\begin{document}

\preprint{arXiv:1207.4073v3 [hep-th]}

\title{The Rotating Black Hole in Renormalizable Quantum Gravity:
The Three-Dimensional Ho\v{r}ava Gravity Case}

\author{Mu-In Park\footnote{E-mail address: muinpark@gmail.com}}

\affiliation{ The Institute of Basic Sciences,
 Kunsan National University, Kunsan, 573-701, Korea }

\begin{abstract}
Recently Ho\v{r}ava proposed a renormalizable quantum gravity,
without the ghost problem, by abandoning Einstein's equal-footing
treatment of space and time through the anisotropic scaling
dimensions.
%, $[t]=-1, [{\bf x}]=-z$ with the dynamical critical
%exponents $(z>1)$.
Since then various interesting aspects, including the exact black
hole solutions have been studied
%In particular, it is rather
%curious that there exist exact black hole solutions even though
%gravitons have different velocity depending on their momenta such as
%the concept of absolute light-cone and so the concept of black hole
%horizon seem to be quite odd in the Lorentz-violating gravity
%theory.
but no {\it rotating} black hole solutions have been found yet,
except some limiting cases. %, and also
%and so there have been some gap in
%Ho\v{r}ava gravity for describing our real black holes in the sky.
%Nevertheless,
%there have been some questions regarding the existence of
%rotating black hole solutions. %for some reasons.
In order to fill the gap, %and clarify this issue,
I consider a simpler three-dimensional set-up with $z=2$
%, where only one UV Lorentz-violating term of
%higher-spatial derivative term is needed, as well as IR
%Lorentz-violating parameters. I
and obtain the exact rotating black hole solution.
% and studied its physical properties.
This solution has a ring curvature singularity inside the outer
horizon, like the four-dimensional Kerr black hole in Einstein
gravity, as well as a curvature singularity at the origin. The usual
mass bound works also here but in a modified form. Moreover, it is
shown that the conventional first law of thermodynamics with the
usual
Hawking temperature and chemical potential % at the outer horizon
does not work, which seems to be the genuine effect of
Lorentz-violating gravity due to lack of the absolute horizon.
\end{abstract}

\pacs{04.20.Jb, 04.20.Dw, 04.60.Kz, 04.60.-m, 04.70.Dy }

\maketitle

\newpage

\section{Introduction}

Recently Ho\v{r}ava proposed a renormalizable gravity theory,
without the ghost (i.e., unitarity) problem, which reduces to
Einstein gravity in IR but with improved UV behaviors, by abandoning
Einstein's equal-footing treatment of space and time through the
anisotropic scaling dimensions, $[t]=-1, [{\bf x}]=-z$ with the
dynamical critical exponents $(z>1)$ \ci{Hora:2008}. Since then
various aspects have been studied, in particular several exact black
hole solutions have been found
% even though the concept of black
%hole horizon is rather curious in the Lorentz-violating gravity
%theory since gravitons have different speed depending on their
%momenta such that the absolute light-cone, which is crucial for
%black hole horizon, is absent but {\it emergent} in IR
\ci{Lu,Cai,Colg,Keha,Park:0905,Cho,Kiri,Capa,Myun:2010}. But no {\it
rotating} black hole solutions have been found yet, except some
limiting cases \ci{Ghod} and so there have been some gap in
Ho\v{r}ava gravity for describing our real black holes in the sky,
which can be even nearly extremal, for example, $c_\ell J/G M^2
>0.98$ in GRS 1915+105 \ci{McCl} for the speed of light $c_\ell$.
%On the other hand, since the mass
%bound condition $G M^2 \geq c_\ell J$
% from the cosmic censorship, i.e., no naked singularity,
%for Kerr black hole in Einstein gravity has the similar role of the
%particle energy bound $E=\sqrt{{\bf p}^2 c_\ell^2+m^2 c_\ell^4} \geq
%|{\bf p}| c_\ell$ from the absolute speed limit $v \leq c_\ell$ for
%the speed of light $c_\ell$, the existence of rotating black hole
%solutions might be questioned due to lack of absolute speed limit of
%gravitons in the Lorentz-violating gravity. \footnote{This question
%has been raised by Gungwon Kang.}

In order to fill the gap, % and clarify the mass bound issue,
in this paper I %first
consider the three-dimensional set-up with $z=2$,
%, where only one UV
%Lorentz-violating term of higher-spatial derivative term $R^2$ is
%required for renormalizability, as well as IR Lorentz-violating
%parameters,
instead of studying the more challenging four-dimensional Ho\v{r}ava
gravity with $z=3$.
%The three-dimensional space-time is simple
%enough generally to enormously reduce the complications in four
%dimensions. However, this space-time is not too simple to get some
%trivial results: For example, there exist black holes, like the BTZ
%black hole for a negative cosmological constant \ci{Bana}, even
%though there are no propagating gravitons in Einstein gravity.
By solving the three coupled non-linear equations for the
three-dimensional $z=2$ Ho\v{r}ava gravity with the general
axisymmetric metric ansatz, I obtain the exact rotating black hole
solution and study its physical properties. This solution has a ring
curvature singularity inside the outer horizon, like the
four-dimensional Kerr black hole in Einstein gravity, as well as a
curvature singularity at the origin. The usual mass bound works also
here but in a modified form. Moreover, it is shown that the
conventional first law of thermodynamics with the usual Hawking
temperature and chemical potential %at the outer horizon
does not work, which seems to be the genuine effect of
Lorentz-violating gravity due to lack of the absolute horizon.

\section{The rotating black hole in three-dimensional Ho\v{r}ava gravity}
%To this ends,
%In order to study the three-dimensional rotating black hole, I start
%by considering
Using the ADM decomposition of the metric
\begin{\eq}
ds^2=-N^2 c_\ell^2 dt^2+g_{ij}\left(dx^i+N^i dt\right)\left(dx^j+N^j
dt\right)\,
\end{\eq}
the three-dimensional renormalizable action with $z=2$
\ci{Soti:2011,Ande}, up to surface terms, %which reads
is given by \footnote{In three-dimensional Lorentz-invariant
space-time, it has been argued that the topologically massive
gravity \ci{Dese:1981} may be renormalizable if suitable
regularization is given \ci{Dese:1990}. But this action, which
violates parity, exists only in the three dimensions and the
renormalizability can not be generalized to four dimensions.
Moreover, recently it has been clarified that the unitarity and
renormalizaton are not compatible in three-dimensional
Lorentz-invariant higher-curvature gravities, which preserves
parity, for general coefficients of the higher-curvature terms
\ci{Mune}, including the new massive gravity case \ci{Berg}.}
\begin{\eq}
I=\frac{1}{\kappa} \int dt d^2 x \sqrt{g} N \left(
K_{ij}K_{ij}-\lambda K^2 +\xi R +\alpha R^2 -2 \Lambda\right),
\label{action}
\end{\eq}
where $\kappa=16 \pi G_3$,
\begin{\eq}
 K_{ij}=\frac{1}{2N}\left(\dot{g}_{ij}-\nabla_i
N_j-\nabla_jN_i\right)\
 \end{\eq}
is the extrinsic curvature, $R$ is the Ricci scalar of the Euclidean
two-geometry, $\lambda,\xi$ are the IR Lorentz-violating parameters,
and $\La$ is the cosmological constant. Note that in two-spatial
dimensions all curvature invariants can be expressed by the Ricci
scalar due to the identities, $R_{ijkl}=(g_{ik} g_{jl}-g_{il}
g_{jk}) R/2,~ R_{ij}= g_{ij} R/2$. Here, I do not consider the terms
which depend on $a_i \equiv \pa_i N/N$ and $\nabla_j a_i$, which can
change the IR as well as UV behaviors a lot from that of
(\ref{action}). Moreover, I do not consider the term of $\nabla ^2
R$ \ci{Soti:2011} either since the qualitative structure of the
solutions I will get is expected to be similar, as in the four
dimensions \ci{Kiri}.

Let me consider now an axially symmetric solution with the metric
ansatz (I adopt the convention of $c_\ell \equiv 1$, hereafter)
\begin{\eq}
ds^2=-N^2(r) dt^2+\f{1}{f(r)} dr^2+r^2\left(d \phi+N^\phi(r)
dt\right)^2.
\end{\eq}
Note that there is no angle $(\phi$) dependance in the metric due to
the circular symmetry in the two-dimensional space even with the
rotation. By substituting the metric ansatz into the action
(\ref{action}), the resulting reduced Lagrangian, after angular
integration, is given by
\begin{\eq}
{\cal{L}}&=&\frac{2\pi}{\kappa}\frac{N}{\sqrt{f}} \left[\f{f r^3
\left({N^{\phi}}^{'} \right)^2}{2 N^2}-\xi f' +\alpha
\f{f'^2}{r}-2\Lambda r \right]\ ,
\end{\eq}
where the prime $(')$ denotes the derivative with respect to $r$.
Note that there is only the $\xi$ dependance but no $\lambda$
dependance in the Lagrangian.
%Here, I obtain the general solution with an arbitrary
%$\La$ and $\xi$.

The equations of motions are
\begin{\eq}
&&-\f{f r^3 ({N^{\phi}}')^2}{2 N^2}-\xi f' +\alpha
\f{f'^2}{r}-2\Lambda r=0,\\
&&\left( \f{\sqrt{f}}{N} r^3 {N^{\phi}}'\right)'=0, \\
&& \left( \f{N}{\sqrt{f}}\right)' \left( 2 \alpha \f{f'}{r}
-\xi\right)+2 \alpha \f{N}{\sqrt{f}} \left(
\f{f''}{r}-\f{f'}{r^2}\right)=0
\end{\eq}
by varying the functions $N$, $N^{\phi}$, and $f$, respectively.

For arbitrary $\alpha$, $\La$ and $\xi$, I obtain the general
solution
\begin{\eq}
f&=&-{\cal M} +\f{b r^2}{2} \left[ 1- \sqrt{a+
\f{c}{r^4}}+\sqrt{\f{c}{r^4}}
ln\left(\sqrt{\f{c}{ar^4}}+\sqrt{1+\f{c}{ar^4}} \right)
\right], \no  \\
\f{N}{\sqrt{f}}&\equiv& W=1-{ln} \sqrt{1+\f{c}{ar^4}} , \no \\
% \label{W_Horava} \\
N^{\phi}&=&-\f{{\cal J}}{2r^2}\left[ 2-{ln} \sqrt{1+\f{c}{a r^4}}
-\sqrt{\f{a r^4}{c}} {arctan}\left(\sqrt{\f{c}{a r^4}}
\right)\right] \label{f_Horava} %\label{N_phi_Horava}
\end{\eq}
with
\begin{\eq}
a=1+\f{8 \alpha \Lambda}{\xi^2},~b=\f{\xi}{2 \alpha},~ c=\f{2 \alpha
{\cal J}^2}{\xi^2}. \label{abc}
\end{\eq}
Here, I have set $W(\infty) \equiv 1,~ N^{\phi}(\infty) \equiv 0$ by
choosing the appropriate coordinate system, without loss of
generality, but they can be conventionally kept as independent
parameters for the analysis of the mass and angular momentum of the
solution. Note that the parameters $a, c$ are restricted to zero or
positive values, i.e., $a, c \geq 0$, or equivalently,
\begin{\eq}
\f{8 \alpha \Lambda}{\xi^2} \geq -1 \label{alpha}
\end{\eq}
and $\alpha \geq 0$, for the real-valued metric functions $f,~N$,
and $N^{\phi}$.

For large $r$  and small $\alpha$, %limit,
the solution expands as
\begin{\eq}
f
%&=&\f{b}{2} (1-\sqrt{a}) r^2-{\cal M} +\f{bc}{4 \sqrt{a}} \f{1}{r^2}
%-\f{c^2}{48 a^{3/2}} \f{1}{r^6} +{\cal O}\left( \f{1}{r^{10}}
%\right)
%\no \\
&=&-\f{\Lambda}{\xi} r^2 \left(1-\f{2 \alpha \Lambda}{\xi^2}\right)
-{\cal M} +\f{{\cal J}^2}{4 r^2 \xi} \left(1-\f{4 \alpha
\Lambda}{\xi^2} \right)  -\f{\alpha {\cal J}^4}{24 \xi^3} \f{1}{r^6}
+{\cal O}(\alpha^2, {r^{-10}} ), \no \\
W
%&=& 1-\f{c}{2 a r^4} +{\cal O} \left( \f{1}{r^8}\right)
&=& 1-\f{\alpha {\cal J}^2}{\xi^2}\f{1}{r^4} +{\cal O} ( \alpha^2,{r^{-8}}),
\no \\
N^{\phi}
%&=&-\f{{\cal J}}{2 r^2} +\f{c {\cal J} }{12 a} \f{1}{r^6} +{\cal O}
%\left(\f{1}{r^{10}} \right)
&=&-\f{{\cal J}}{2 r^2} +\f{\alpha {\cal J}^3}{6 \xi^2} \f{1}{r^6}
+{\cal O} (\alpha^2,{r^{-10}} ).
\end{\eq}
 It is easy to check that, in the limit of $\alpha
\rightarrow 0$, the solution reduces to the BTZ black hole solution
(with $\xi=1$) \ci{Bana}
\begin{\eq}
N^2_{BTZ}=f_{BTZ}=-\f{\Lambda}{\xi} r^2 -{\cal M}+\f{{\cal J}^2}{4
r^2 \xi},~ N^{\phi}_{BTZ}=-\f{{\cal J}}{2 r^2}.
\end{\eq}
%and so the solution (\ref{f_Horava}) %-(\ref{N_phi_Horava}) are
%is the Ho\v{r}ava-gravity generalization of the BTZ black hole.

The non-vanishing curvature invariants are
\begin{\eq}
R&=&-\f{f'}{r}=-b \left(1-\sqrt{a+ \f{c}{r^4}} \right),  \label{R}\\
K^{ij} K_{ij} &=&\f{r^2}{2 W^2} \left( {N^{\phi}}^{'} \right)^2 \no
\\ &=&\f{{\cal J}^2}{2 r^4} \f{\left(ln\sqrt{1+ \f{c}{a
r^4}}\right)^2}{\left(1-ln\sqrt{1+ \f{c}{a r^4}}\right)^2}
\label{KK}
\end{\eq}
and (\ref{KK}) shows a ring curvature singularity %in $K^{ij} K_{ij}$
when $W=0$, i.e., at
\begin{\eq}
r_{ring}=\left( \f{c}{a (e^2-1)}\right)^{1/4} \approx \left( 0.1565~
\f{c}{a }\right)^{1/4} \label{ring}
\end{\eq}
for the rotating solution, as well as a curvature singularity at
$r=0$ in both $R$ and $K^{ij} K_{ij}$. %the origin.

Note that the existence of a ring singularity is analogous to the
four-dimensional Kerr black hole case but the singularity at $r=0$
is not. For the BTZ black hole in Einstein gravity $(\alpha=0,~ \xi
= 1)$, the curvature singularity at $r=0$ in $R$ is canceled by
$K^{ij} K_{ij}-K^2$ and the remainders, which become the boundary
terms in the action, in the (covariant) three curvature scalar
$R^{(3)}$, resulting the finite value: $R^{(3)}=R+K^{ij}
K_{ij}-K^2-f'/r-f''=6 \Lambda$. This means that the curvature
singularity at $r=0$ in $R$ is the artifact of the time-foliation
and not the physical singularity in Einstein gravity. But for the
general
solutions (\ref{f_Horava}), %-(\ref{N_phi_Horava}),
the curvature singularities at $r=0$ %the origin
in (\ref{R}) and (\ref{KK}) are covariant in the foliation
preserving diffeomorphism such that they are physical
singularities.
%\footnote{The singularities can not be removed even in
%the combination of $R^{(3)}$, either.}

For asymptotically AdS, i.e., $\Lambda <0$ \footnote{For
asymptotically dS, i.e., $\Lambda >0$, the solution (\ref{f_Horava})
has the cosmological horizon as a generalization of $KdS_3$ in
Einstein gravity \ci{Park:1998}. And also, for asymptotically flat,
i.e., $\Lambda=0$, (\ref{f_Horava}) has no horizon either. However,
I will not consider these geometries here since the curvature
singularities are then naked.}, the solution (\ref{f_Horava})
%,(\ref{W_Horava}) have
has two horizons generally where $f$ and $N$ vanish simultaneously,
i.e., the apparent and Killing horizons coincide, and the Hawking
temperature for the outer horizon $r_+$ is given by

\begin{\eq}
T_+&=&\f{ \hbar (W f')|_{r_+}}{4 \pi} \no \\
&=&\f{\hbar}{4 \pi} b r_+\left(1-\sqrt{a+\f{c}{r_+^4}} \right)
\left(1-ln \sqrt{1+\f{c}{ar_+^4}}\right) \label{Temp}
\end{\eq}
from the regularity of the horizon in the Euclidean space-time, as
usual. There is another Killing horizon when $W=0$, i.e.,
$N=W\sqrt{f}=0$ with $f \neq 0$, at $r=r_{ring}$ but this is not the
event horizon since one can escape from (or reach to) the horizon in
a finite time.\footnote{
 This may be compared with the non-commutative BTZ
black hole case \ci{Park:2007}, where there is no coincidence point
of the apparent and the Killing horizons and the smeared region is
formed between them.}

\begin{figure}
\includegraphics[width=8cm,keepaspectratio]{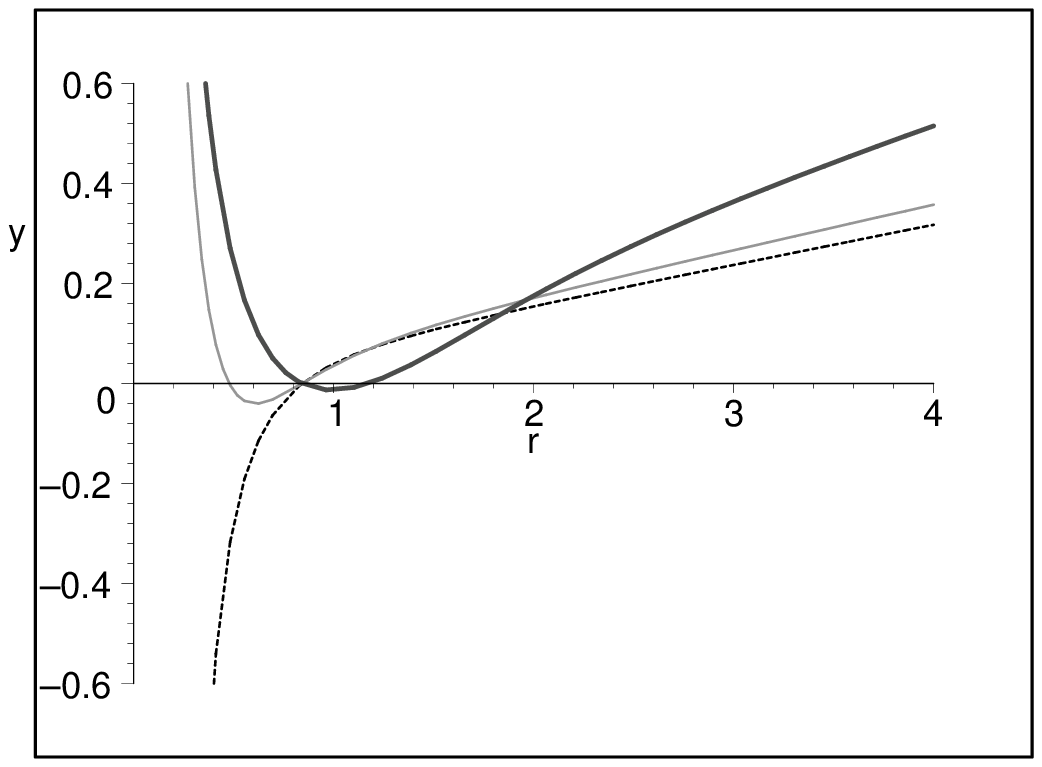}
\includegraphics[width=8cm,keepaspectratio]{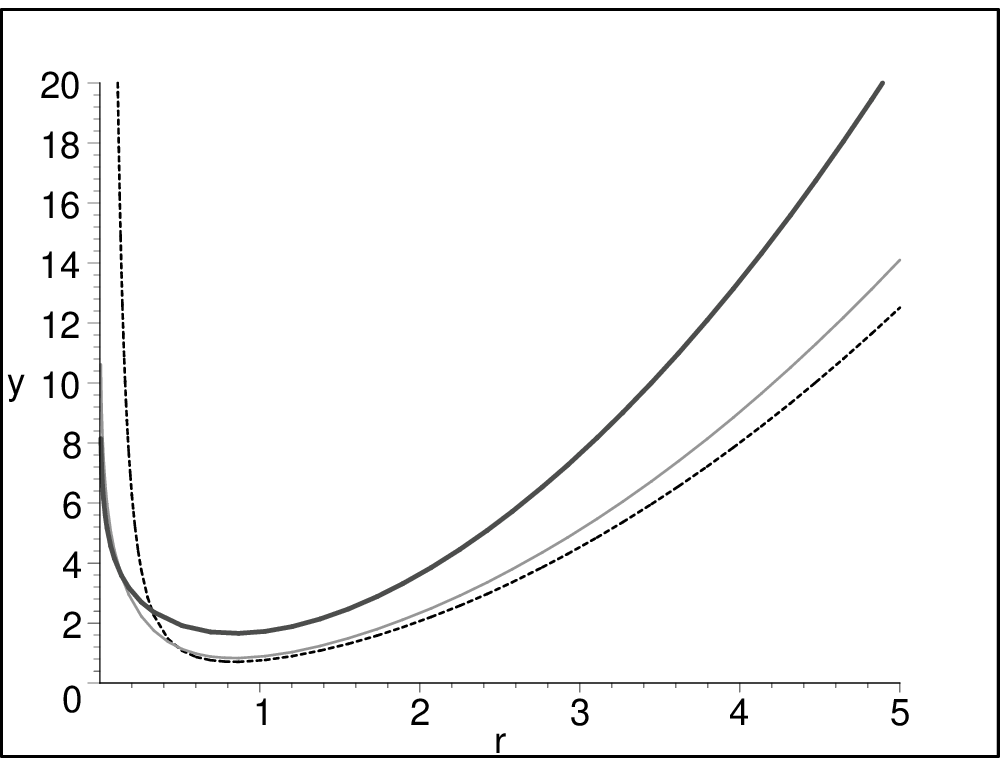}
\caption{Plots of $T_+$ (left) and ${\cal M}$ (right) vs.
%horizon radius
$r_+$ for AdS space. The two solid curves represent the
three-dimensional rotating Ho\v{r}ava black holes for different
Lorentz-violating higher-derivative coupling $\alpha=0.24,~0.1$ for
the dark and bright curves, respectively, in comparison with the BTZ
case ($\alpha=0$) in the dotted curve. Here, I have considered
$\xi=1,\Lambda=-0.5,{\cal J}=1$, and $\hbar \equiv 1$.
}\label{fig:Temp_AdS}
\end{figure}
In Fig.\ref{fig:Temp_AdS} (left), the temperature $T_+$ vs. the
outer horizon radius $r_+$ is plotted. For non-vanishing $c$, i.e.,
$\alpha, {\cal J} \neq 0$, there are two instances of the vanishing
temperature:

(a) The first case is the usual extremal black hole limit, where the
inner horizon $r_+$ meets the outer horizon $r_+$ at
\begin{\eq}
r_{+}^{*}=\left( \f{c}{1-a}\right)^{1/4}
\end{\eq}
and the integration constant
\begin{\eq}
{\cal M} =\f{b r_+^2}{2} \left[ 1- \sqrt{a+
\f{c}{r_+^4}}+\sqrt{\f{c}{r^4}}
{ln}\left(\sqrt{\f{c}{ar_+^4}}+\sqrt{1+\f{c}{ar_+^4}} \right)
 \right]
\end{\eq}
gets the minimum (Fig.\ref{fig:Temp_AdS} (right)). This is the
ground state in the usual black hole system and the outer horizon
can not be smaller than $r_+^*$;
%The extremal
%radius $r_+^*$ is the Cauchy horizon and so continuation to the
%region of $r_+ < r_+^*$ does not make any sense to outside observer;
$T_+<0$ for $r_+ < r_+^*$ and this reflects a pathology of the
region ( for some related discussions, see Ref. \ci{Park:2006}).

(b) The second case is the instance when $W|_{r_+}=0$, i.e., when
the outer horizon $r_+$ meets the ring curvature singularity
$r_{ring}$ of (\ref{ring}). If $\alpha$ is small enough so that
$a>1/e^2$, i.e., $-8 \alpha \Lambda/\xi^2 < 1-1/e^2$, then this
instance does not really occur since the outer horizon is always
larger than the radius of the ring curvature singularity $r_{ring}$,
i.e., $r_{ring}<r_+^* \leq r_+$. In this case the zero temperature
is arrived when $r_+$ meets $r_-$ at $r_+^*$ {\it before} meets
$r_{ring}$. This shows that the ring curvature singularity is safely
protected by the outer horizon for the small $\alpha$. However, if
$\alpha$ is not so small so that $a\leq 1/e^2$, i.e., $-8 \alpha
\Lambda/\xi^2 \geq 1-1/e^2$, then there is the chance when $r_+$
meets $r_{ring}$ from outside, i.e., $r_{+} \geq r_{ring}$. But even
in this case the ring singularity would not be naked since the zero
temperature, i.e., the ground state is arrived by merging $r_+ \ra
r_{ring}$, and $r_+$ can not be smaller than $r_{ring}$: There is
the ring singularity ``on'' the horizon, but this does not affect
the outer region $(> r_+)$ by the definition of the event horizon.
In this case the second instance of the zero temperature is arrived
before reaching the extremal black hole, except the case $a=1/e^2$,
where the extremal and ring singularity radius are degenerate,
$r_+^*=r_{ring}$.
%\footnote{If $r_+$ becomes smaller than $r_{ring}$
%such that the ring singularity is naked, the Hawking temperature
%becomes negative which seems to imply the semi-classical instability
%of the naked space-time.}

The ergo-region is defined by $g_{tt}=-N^2 +r^2 (N^{\phi})^2 \geq 0$
with its boundary at
\begin{\eq}
r_{erg}=\left.\f{\sqrt{f} |W|}{|N^{\phi}|}\right|_{r=r_{erg}}
\label{r_erg}
\end{\eq}
and this region is outside of the outer horizon $r_+$, i.e., $r_+
\leq r_{erg}$ since $f(r_{erg})=g_{rr}(r_{erg})\geq 0$ is required
by (\ref{r_erg}). Here, the properties $W(r_{erg})>0$ from $r_{erg}
\geq r_+
> r_{ring}$ and $N^{\phi}>0$ are used.

\begin{figure}
\includegraphics[width=10cm,keepaspectratio]{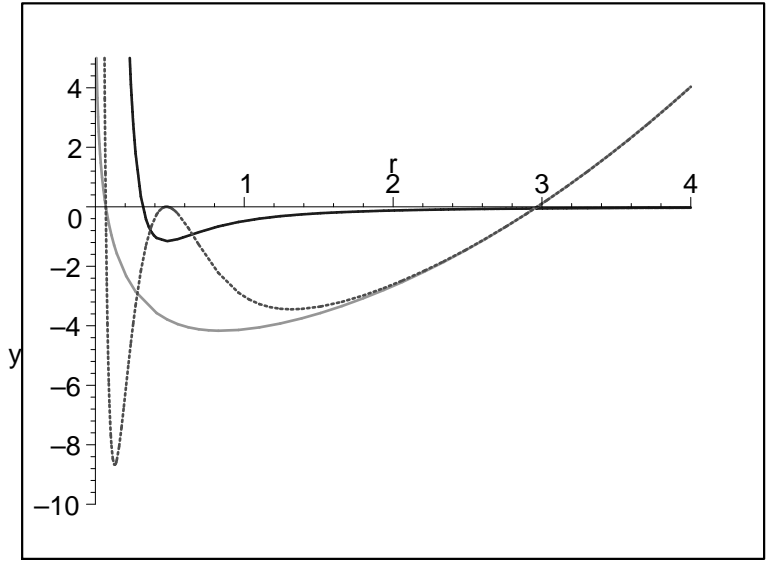}
\caption{Plots of $f(r)$ (bright solid), $N^2(r)=W^2 f$ (dotted),
$N^{\phi}(r)$ (dark) curves for AdS space ($\xi=1, \Lambda=-0.5,
{\cal M}=5, {\cal J}=1, \alpha=0.1$). In addition to the two
horizons $r_-, r_+$ which are solutions of $f=0,~N^2=0$,
simultaneously, $N^2$ has one more additional solution of $N^2=W^2
f=0~(f \neq 0)$ at $r_{ring}$ where the ring singularity is located,
between $r_-$ and $r_+$. There is also the counter-rotation
($N^{\phi}>0$) for $r<r_{count}$ and the turning point
(${N^{\phi}}'=0$) is at $r_{ring}$. } \label{fig:N_f}
\end{figure}

Another peculiar property of the general solution is that there is
the counter-rotating region inside the outer horizon $r_+$
(Fig.\ref{fig:N_f}).
It is interesting to note that the turning point of $N^{\phi}$,
i.e., ${N^{\phi}}'=0$ is at the location of the ring singularity
$r=r_{ring}$ from ${N^{\phi}}'=W {\cal J}/r^3=0$ and the
counter-rotating region starts at $r_{count}=(\eta c/a)^{1/4}~ (<
r_{ring}<r_+)$ with $\eta\approx0.0308$ which solves
$N^{\phi}=0~({\cal J}\neq 0)$.
%\footnote{This seems to imply that the
%ring curvature singularity carries the intrinsic angular momentum in
%the counter-rotating direction.}

\section{The Unusual thermodynamics}
The thermodynamics of Lorentz-violating black holes has not been
well established yet \footnote{But, see Ref. \ci{Myun} for the black
hole thermodynamics of non-rotating Ho\v{r}ava black holes in four
dimensions.}. In order to study this subject, I start by computing
the conserved mass and angular momentum of the rotating black
solution
(\ref{f_Horava}). %-(\ref{N_phi_Horava}).
To this ends, let me consider the variation of the total action
$I_{total}=I+B$ with boundary terms $B$ at space-like infinity such
that the boundary variation $(\de I)(\infty)$ is canceled by $\de B$
and there remain only the bulk terms in $\de I_{total}$ which vanish
when the equations of
motions hold. % , i.e., differentiable.
Then for the class of fields that approach our solution
(\ref{f_Horava}) %-(\ref{N_phi_Horava})
at infinity, one finds %that
\begin{\eq}
B=(t_2-t_1) (-W(\infty) M+N^{\phi} (\infty) J),
\end{\eq}
which defines the canonical mass and angular momentum
\begin{\eq}
M=\f{2 \pi \xi \sqrt{a}}{\kappa} {\cal M}, ~J=\f{2 \pi \xi }{\kappa}
{\cal J},
\end{\eq}
as the conjugates to the asymptotic displacements $N(\infty)$ and
$N^{\phi}(\infty)$, respectively, when kept as independent
parameters.

In order that the curvature singularities are not naked, i.e.,
satisfying the cosmic censorship, one needs the mass bound condition
\begin{\eq}
M \geq \chi (x) J \sqrt{-\Lambda}/|\xi|~  \label{mass_bound}
\end{\eq}
with the monotonic function
\begin{\eq}
\chi(x)=\sqrt{x^2-1} ~{ln} \left(
\f{1}{\sqrt{x^2-1}}+\f{1}{\sqrt{1-x^{-2}}}\right)
\end{\eq}
which can vary in $[0,~1]$ as $ x^2 \equiv \xi^2/(-8 \Lambda
\alpha)$ varies in $[1,~\infty]$. In the BTZ limit,
$x^2=\infty,~\xi=1$, one has the usual mass bound ($\chi=1$)
\begin{\eq}
M \geq J \sqrt{-\Lambda},
\end{\eq}
but even for the other more general classes of $1 \leq \chi (x) <
\infty$ so that $0 \leq \chi <1$, the mass bound still works for
each theory parameterized by $x$, but in a modified form.

Now in order to study the first law of black hole thermodynamics,
let me consider the variation of the mass $M$ as a function of $J$
and $r_+$,
\begin{\eq}
dM =A dJ +B d r_+ \label{first:0}
\end{\eq}
with
\begin{\eq}
A&=&\f{\kappa J}{4 \pi \xi^2} \sqrt{\f{a}{c}}~
{ln}\left(\sqrt{\f{c}{ar_+^4}}+\sqrt{1+\f{c}{ar_+^4}} \right), \no \\
B&=&\f{\pi \xi^2}{\kappa \alpha} r_+  \sqrt{a} \left(
1-\sqrt{a+\f{c}{r_+^4}} \right) \label{A,B}.
\end{\eq}

\begin{figure}
\includegraphics[width=5cm,keepaspectratio]{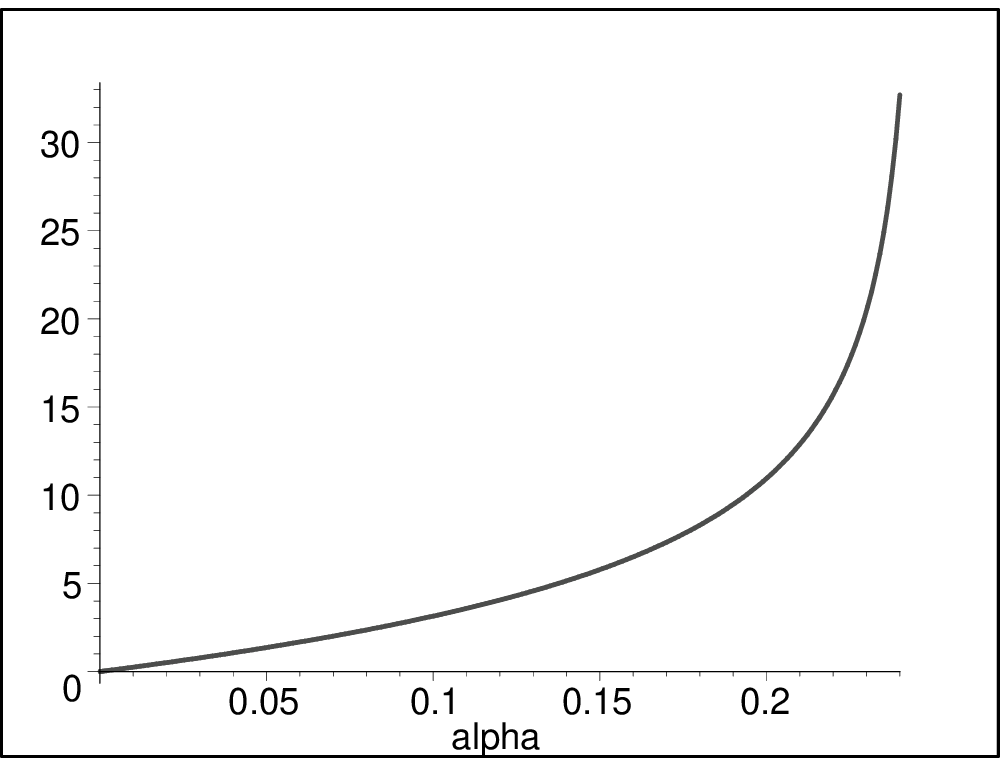}
\includegraphics[width=5cm,keepaspectratio]{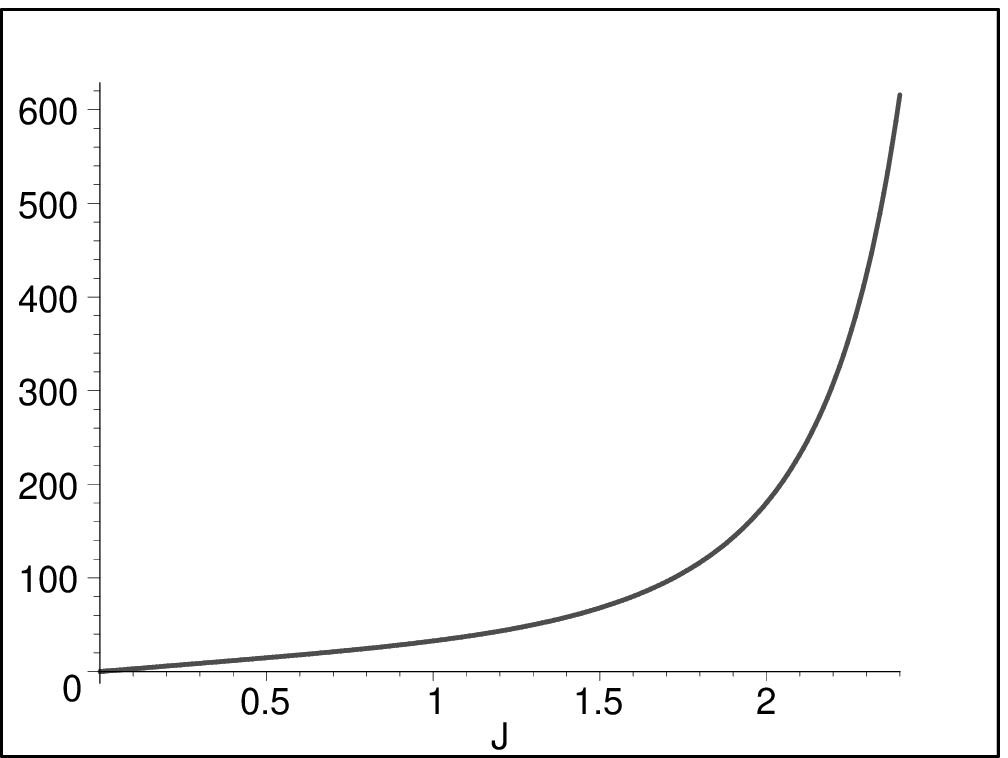}
\includegraphics[width=5cm,keepaspectratio]{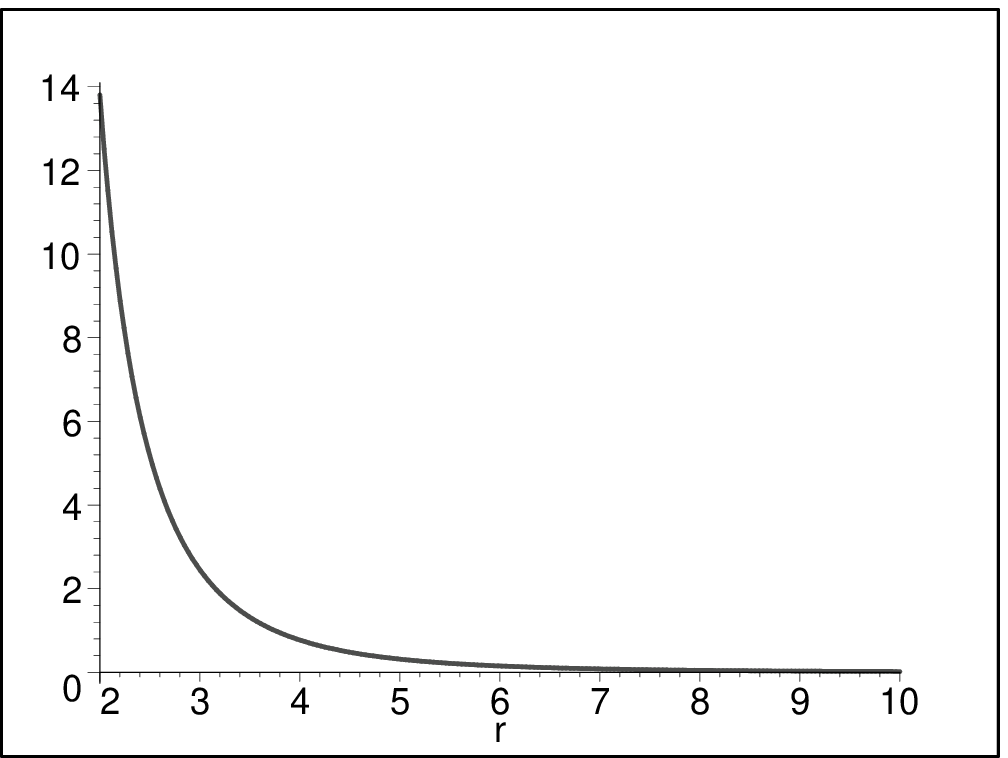}
\caption{Plots of $\partial_{J}
\partial_{r_+} S- \partial_{r_+} \partial_{J} S $ vs. $\alpha$ (left),
${\cal J}$ (middle) and $r_+$ (right)
 for AdS space
($\xi=1,\Lambda=-0.5,{\cal J}=1,\kappa=1,\alpha=0.24$ (middle,
right), $r_+=2$ (left, middle)). The infinite barriers in the left
and middle curves are due to  $\alpha \leq -\xi^2/8 \Lambda$
(\ref{alpha}) and $M \leq \chi(x) J \sqrt{-\Lambda}/|\xi| $
(\ref{mass_bound}). }\label{fig:ddS}
\end{figure}

Then, in order to see whether the first law of thermodynamics in the
conventional form
\begin{\eq}
d M=T_+ d S+\Omega_+ dJ \label{first:1}
\end{\eq}
works with the usual Hawking temperature $T_+$ of (\ref{Temp}) and
the chemical potential $\Omega_+=-N^{\phi}|_+$, let me define the
black hole entropy function $S$ with
\begin{\eq}
dS \equiv \partial_{r_+} S~ d r_+ +\partial_{J} S ~dJ
\end{\eq}
as a function of $r_+$ and $J$. Then, from (\ref{first:0}),
(\ref{A,B}), and (\ref{first:1}), one can find
\begin{\eq}
\partial_{r_+} S=\f{B}{T_+},~
\partial_{J} S=\f{\alpha \kappa^2 J}{ \pi^2 \xi^4 T_+} \left(
A-\Omega_+\left( \f{2 \pi}{\kappa} \right)^2 \f{\xi^3}{\alpha J}
\right)
\end{\eq}
but %one can check that
%the entropy is not integrable, i.e.,
$\partial_{J}
\partial_{r_+} S- \partial_{r_+} \partial_{J} S \neq 0$,
for arbitrary non-vanishing $\alpha$, ${\cal J}$, and finite $r_+$.
The lengthy result for the non-integrability is not so impressive to
be shown here but,
%this paper. Rather,
in order to grasp how the Lorentz violation and the angular momentum
affect the non-integrability, I show %only
its
%whose
leading term %in the analytic form
%is given by
\begin{\eq}
\partial_{J}
\partial_{r_+} S- \partial_{r_+} \partial_{J} S=\f{16 \pi^2 {\cal J
}}{\kappa r_+^4} ~\alpha +{\cal O}(\alpha^2).
\end{\eq}
%but
and the full results in the numerical plots (Fig. \ref{fig:ddS}).
These results show that the entropy is not integrable by the
non-relativistic higher curvature corrections ($\alpha \neq 0$) for
the rotating and
finite % ($r_+ < \infty$)
black holes. The infinite barriers at $\alpha$ and $J$ are due to
$\alpha \leq -\xi^2/8 \Lambda$ (\ref{alpha}) and $M \leq  \chi(x) J
\sqrt{-\Lambda}/|\xi|$ (\ref{mass_bound}). This proves that the
entropy can {\it not} be defined in the conventional form of the
first law of thermodynamics with the usual Hawking temperature and
chemical potential.

\section{Discussion}

In conclusion, I have obtained the rotating black hole solution in
the three-dimensional Ho\v{r}ava gravity where the Lorentz symmetry
is broken by the higher-spatial derivatives in UV. Here, it is
remarkable that the existence of the rotating black hole does not
depend much on the existence nor the momentum dependance of speed,
i.e., no absolute speed limit, of gravitons. Actually, in our case
there would be no graviton mode at the linear
perturbation\footnote{This is in contrast to the extended Ho\v{r}ava
gravity \ci{Soti:2011} or the projectable model
\ci{Ande} where there is one scalar graviton mode.} %order
from the similar analysis in four-dimensional Ho\v{r}ava gravity
since the calculation is not sensitive to the dimensionality of
space \ci{Keha,Park:0910}. The status of its full, non-linear
analysis is still unclear and needs more elaborative works with some
ingenious separation of the genuine constraints, which being left as
a further work.

And %as much remarkable as the existence of black hole solutions in
%the Lorentz-violating gravity,
I have also shown that the mass bound condition still works in the
new solution, analogous to the mass bound $G M^2 \geq c_\ell J$
% from the cosmic censorship, i.e., no naked singularity,
for Kerr black hole in Einstein gravity.
%, which the similar role of
%the particle energy bound $E=\sqrt{{\bf p}^2 c_\ell^2+m^2 c_\ell^4}
%\geq |{\bf p}| c_\ell$ from the absolute speed limit $v \leq c_\ell$
%for the speed of light $c_\ell$,, which has the similar role, in
%Einstein gravity, as the particle energy bound from the absolute
%speed limit $c_{\ell}$
%in relativistic mechanics.
However, I have shown that the first law
of thermodynamics can not be written in the conventional form with
the usual Hawking temperature and the chemical potential such that
the entropy function can not be defined for the generically rotating
and Lorentz-violating black holes. The existence of Hawking
temperature implies the Hawking radiation and this can be proved
quite generally without knowing much details of the solutions (see
for example Ref. \ci{Angh}).

So, we have the black holes which generate the Hawking radiation but
without the black hole entropy. Actually the notion of ``Hawking
radiation without black hole entropy'' has been studied in the
context of analogue black holes \ci{Viss}, previously. In our case,
this seems to be a genuine effect of the Lorentz-violating gravity
due to lack of the absolute horizon which can leak the information
depending on the matter's momentum scale. This may be compared with
other Lorentz-violating black holes, called Lifshitz black holes,
where the first law of thermodynamics does not hold for a generic
member of a class of black holes \ci{Deve}. %Finally,
The study of rotating black holes in four-dimensional Ho\v{r}ava
gravity and their black hole thermodynamics would be quite a
challenging problem.
%which needs to solve five coupled equations, instead of three in our case.

As a possible resolution for the failure of the usual black hole
thermodynamics, one might try to consider the first law of
thermodynamics in the form of $dM=\tilde{T} dS+\Om_+ dJ$ with an
unusual ``temperature'' function $\tilde{T}=\tilde{T}(r_+,J)$ and
its associated entropy function $S=S(r_+,J)$, instead of the
standard one (\ref{first:1}). However, the usual interpretations of
$\tilde{T}$ and $S$ as the thermal temperature of Hawking radiation
and the black hole entropy, respectively, need to be justified.
%On the other hand,
For non-rotating black holes, or more generally one-parameter family
of black hole solutions, one can {\it always} consider the standard
first law of thermodynamics $dM=T_+ dS$ with the appropriate entropy
function $S=S(r_+)$ \ci{Cai}. In our three dimensional case, one can
easily find $T_+=\hbar b r_+(1-\sqrt{a})/4 \pi$, $ S=2 \pi r_+
\xi/4G \hbar$ with the black hole horizon $r_+=(2 {\cal M}/b
(1-\sqrt{a}))^{1/2}$ and this becomes the usual black hole entropy
in three dimensions \ci{Bana} with $\xi =1$, i.e., no Lorentz
violation in IR. Here, it is interesting to note that the UV Lorentz
violation parameter $\alpha$ does not affect the usual entropy
formula but affect only the value of $r_+$ through the parameters
$a$ and $b$ from (\ref{abc}), in contrast to four-dimensional black
holes \ci{Cai,Myun}, where UV Lorentz violation terms produce
logarithmic corrections to the entropy formula. On the other hand,
the modification of the entropy from the usual area ({\it
perimeter}, in our case) law comes from IR Lorentz violation
parameter $\xi$ but a thermodynamic interpretation of the modified
entropy is not obvious.

\section*{Acknowledgments}

I would like to thank Gungwon Kang for giving some inspiration. This
work was supported by the Korea Research Foundation Grant funded by
Korea Government(MOEHRD) (KRF-2010-359-C00009).

%%%%%%%%%% References %%%%%%%%%%%%%%%%%%%%%%%%%
\newcommand{\J}[4]{#1 {\bf #2} #3 (#4)}
\newcommand{\andJ}[3]{{\bf #1} (#2) #3}
\newcommand{\AP}{Ann. Phys. (N.Y.)}
\newcommand{\MPL}{Mod. Phys. Lett.}
\newcommand{\NP}{Nucl. Phys.}
\newcommand{\PL}{Phys. Lett.}
\newcommand{\PR}{Phys. Rev. D}
\newcommand{\PRL}{Phys. Rev. Lett.}
\newcommand{\PTP}{Prog. Theor. Phys.}
\newcommand{\hep}[1]{ hep-th/{#1}}
\newcommand{\hepp}[1]{ hep-ph/{#1}}
\newcommand{\hepg}[1]{ gr-qc/{#1}}
\newcommand{\bi}{ \bibitem}
%%%%%%%%%%%%%%%%%%%%%%%%%%%%%%%%%%%%%%%%%%%%%%%


\begin{thebibliography}{999}

\bibitem{Hora:2008}
  P.~Ho\v{r}ava,
  %``Membranes at Quantum Criticality,''
  JHEP {\bf 0903}, 020 (2009) [arXiv:0812.4287 [hep-th]];
  %%CITATION = JHEPA,0903,020;%%
%\bibitem{Hora} P.~Ho\v{r}ava,
  %``Quantum Gravity at a Lifshitz Point,''
  Phys.\ Rev.\  D {\bf 79}, 084008 (2009)
  [arXiv:0901.3775 [hep-th]].

\bibitem{Lu}
  H.~Lu, J.~Mei and C.~N.~Pope,
  %``Solutions to Horava Gravity,''
Phys.\ Rev.\ Lett.\  {\bf 103}, 091301 (2009)
  [arXiv:0904.1595 [hep-th]].
  %%CITATION = ARXIV:0904.1595;%%

%\bibitem{Nast}
%  H.~Nastase,
%  ``On IR solutions in Horava gravity theories,''
%  arXiv:0904.3604 [hep-th].

\bibitem{Cai}
  R.~G.~Cai, L.~M.~Cao and N.~Ohta,
  %``Topological Black Holes in Horava-Lifshitz Gravity,''
Phys.\ Rev.\ D {\bf 80}, 024003 (2009)
  [arXiv:0904.3670 [hep-th]].

\bibitem{Colg}
  E.~O.~Colgain and H.~Yavartanoo,
  %``Dyonic solution of Horava-Lifshitz Gravity,''
 JHEP {\bf 0908}, 021 (2009)
  [arXiv:0904.4357 [hep-th]]; S.~-S.~Kim, T.~Kim and Y.~Kim,
  %``Surplus Solid Angle: Toward Astrophysical Test of
 %Horava-Lifshitz Gravity,''
 Phys.\ Rev.\ D {\bf 80}, 124002 (2009)
 [arXiv:0907.3093 [hep-th]]; E.~Gruss,
  %``Black Holes in Horava Gravity with Higher Derivative Magnetic Terms,''
  Class.\ Quant.\ Grav.\  {\bf 28}, 085007 (2011)  [arXiv:1005.1353 [hep-th]].

\bibitem{Keha}
  A.~Kehagias and K.~Sfetsos,
 % ``The black hole and FRW geometries of non-relativistic gravity,''
Phys.\ Lett.\ B {\bf 678}, 123 (2009)
  [arXiv:0905.0477 [hep-th]].

\bibitem{Park:0905}
  M.~-I.~Park,
  %``The Black Hole and Cosmological Solutions in IR modified Ho\v{r}ava Gravity,''
  JHEP {\bf 0909}, 123 (2009)
    [arXiv:0905.4480 [hep-th]].

\bi{Cho}
  I.~Cho and G.~Kang,
 % ``Four dimensional string solutions in Ho\v{r}ava-Lifshitz gravity,''
  JHEP {\bf 1007}, 034 (2010)
    [arXiv:0909.3065 [hep-th]];
    A.~N.~Aliev and C.~Senturk,
  %``Black Strings in Ho\v{r}ava-Lifshitz Gravity,''
  Phys.\ Rev.\ D {\bf 84}, 044010 (2011)  [arXiv:1106.0024 [hep-th]].

\bi{Kiri}
  E.~B.~Kiritsis and G.~Kofinas,
 %``On Ho\v{r}ava-Lifshitz 'Black Holes',''
  JHEP {\bf 1001}, 122 (2010)
  [arXiv:0910.5487 [hep-th]];
  G.~Koutsoumbas and P.~Pasipoularides,
  %``Black hole solutions in Horava-Lifshitz Gravity with cubic terms,''
  Phys.\ Rev.\ D {\bf 82}, 044046 (2010) [arXiv:1006.3199 [hep-th]];
  G.~Koutsoumbas, E.~Papantonopoulos, P.~Pasipoularides and M.~Tsoukalas,
  %``Black Hole Solutions in 5D Horava-Lifshitz Gravity,''
  Phys.\ Rev.\ D {\bf 81}, 124014 (2010)  [arXiv:1004.2289 [hep-th]].

\bibitem{Capa}
  D.~Capasso and A.~P.~Polychronakos,
  %``General static spherically symmetric solutions in Horava gravity,''
  Phys.\ Rev.\ D {\bf 81}, 084009 (2010)  [arXiv:0911.1535 [hep-th]].

\bibitem{Myun:2010}
  Y.~S.~Myung,
  %``Lifshitz black holes in the Ho\v{r}ava-Lifshitz gravity,''
  Phys.\ Lett.\ B {\bf 690}, 534 (2010)  [arXiv:1002.4448 [hep-th]].

\bibitem{Ghod}
  A.~Ghodsi,
  %``Toroidal solutions in Ho\v{r}ava Gravity,''
  Int.\ J.\ Mod.\ Phys.\ A {\bf 26}, 925 (2011)
  [arXiv:0905.0836 [hep-th]];
%\bibitem{Ghodsi:2009zi}
  A.~Ghodsi and E.~Hatefi,
  %``Extremal rotating solutions in Ho\v{r}ava Gravity,''
  Phys.\ Rev.\ D {\bf 81}, 044016 (2010)
  [arXiv:0906.1237 [hep-th]];
    H.~W.~Lee, Y.~-W.~Kim and Y.~S.~Myung,
  %``Slowly rotating black holes in the Ho\v{r}ava-Lifshitz gravity,''
  Eur.\ Phys.\ J.\ C {\bf 70}, 367 (2010)
   [arXiv:1008.2243 [hep-th]];
  %\bibitem{Aliev:2010eg}
  A.~N.~Aliev and C.~Senturk,
  %``Slowly Rotating Black Hole Solutions to Ho\v{r}ava-Lifshitz Gravity,''
  Phys.\ Rev.\ D {\bf 82}, 104016 (2010)
    [arXiv:1008.4848 [hep-th]].

\bibitem{McCl}
  J.~E.~McClintock, %{\it et. al.},
  R.~Shafee, R.~Narayan, R.~A.~Remillard, S.~W.~Davis and L.~-X.~Li,
  %``The Spin of the Near-Extreme Kerr Black Hole GRS 1915+105,''
  Astrophys.\ J.\  {\bf 652}, 518 (2006)
  [astro-ph/0606076].

\bi{Soti:2011}
  T.~P.~Sotiriou, M.~Visser and S.~Weinfurtner,
  %``Lower-dimensional Ho\v{r}ava-Lifshitz gravity,''
 Phys.\ Rev.\ D {\bf 83}, 124021 (2011)
  [arXiv:1103.3013 [hep-th]].

\bi{Ande}
  C.~Anderson, %{\it et. al.},
  S.~J.~Carlip, J.~H.~Cooperman, P.~Ho\v{r}ava, R.~K.~Kommu and P.~R.~Zulkowski,
  %``Quantizing Ho\v{r}ava-Lifshitz Gravity via Causal Dynamical Triangulations,''
  Phys.\ Rev.\ D {\bf 85}, 044027 (2012)
    [arXiv:1111.6634 [hep-th]].

\bibitem{Dese:1981}
  S.~Deser, R.~Jackiw and S.~Templeton,
%``Three-Dimensional Massive Gauge Theories,'' Phys.\ Rev.\ Lett.\
{\bf 48}, 975 (1982); %``Topologically Massive Gauge Theories,''
  Annals Phys.\  {\bf 140}, 372 (1982) [Erratum-ibid.\  {\bf 185}, 406 (1988)]  [Annals Phys.\  {\bf 185},
406 (1988)] [Annals Phys.\  {\bf 281}, 409 (2000)].

\bibitem{Dese:1990}
  S.~Deser and Z.~Yang,
%  ``Is Topologically Massive Gravity Renormalizable?,''
  Class.\ Quant.\ Grav.\  {\bf 7}, 1603 (1990).

\bibitem{Mune}
  K.~Muneyuki and N.~Ohta,
 % ``Unitarity versus Renormalizability of Higher Derivative Gravity in 3D,''
  Phys.\ Rev.\ D {\bf 85}, 101501 (2012)
    [arXiv:1201.2058 [hep-th]].

\bibitem{Berg}
  E.~A.~Bergshoeff, O.~Hohm and P.~K.~Townsend,
  %``Massive Gravity in Three Dimensions,''
 Phys.\ Rev.\ Lett.\  {\bf 102}, 201301 (2009)
   [arXiv:0901.1766 [hep-th]].

\bibitem{Bana}
  M.~Banados, C.~Teitelboim and J.~Zanelli,
  %``The Black hole in three-dimensional space-time,''
  Phys.\ Rev.\ Lett.\  {\bf 69}, 1849 (1992).
    [hep-th/9204099].

\bibitem{Park:1998}
  M.~-I.~Park,
  %``Statistical entropy of three-dimensional Kerr-de Sitter space,''
    Phys.\ Lett.\ B {\bf 440}, 275 (1998)
      [hep-th/9806119].

\bibitem{Park:2007}
  H.~-C.~Kim, %{\it et. al. },
  M.~-I.~Park, C.~Rim and J.~H.~Yee,
  %``Smeared BTZ Black Hole from Space Noncommutativity,''
  JHEP {\bf 0810}, 060 (2008)
    [arXiv:0710.1362 [hep-th]].

\bibitem{Park:2006}
M.~-I.~Park,
  %``Thermodynamics of Exotic Black Holes, Negative Temperature, and
  %Bekenstein-Hawking Entropy,''
  Phys.\ Lett.\  B {\bf 647}, 472 (2007)
   [arXiv:hep-th/0602114];
%M.~I.~Park,
  %``BTZ black hole with gravitational Chern-Simons: Thermodynamics and
  %statistical entropy,''
  Phys.\ Rev.\  D {\bf 77}, 026011 (2008)
  [arXiv:hep-th/0608165];
%M.~I.~Park,
  %``BTZ black hole with higher derivatives, the second law of  thermodynamics,
  %and statistical entropy,''
  Phys.\ Rev.\  D {\bf 77}, 126012 (2008)
  [arXiv:hep-th/0609027];
%  M.~I.~Park,
  %``Can Hawking temperatures be negative?,''
  Phys.\ Lett.\  B {\bf 663}, 259 (2008)
   [arXiv:hep-th/0610140];
%M.~I.~Park,
  %``Thoughts on the area theorem,''
  Class.\ Quant.\ Grav.\  {\bf 25}, 095013 (2008)
 [arXiv:hep-th/0611048].

\bibitem{Myun}
Y.~S.~Myung and Y.~W.~Kim,
 %``Thermodynamics of Ho\v{r}ava-Lifshitz black holes,''
 Eur.\ Phys.\ J.\ C {\bf 68}, 265 (2010)
 [arXiv:0905.0179 [hep-th]];
  R.~G.~Cai, L.~M.~Cao and N.~Ohta,
  %``Thermodynamics of Black Holes in Ho\v{r}ava-Lifshitz Gravity,''
  Phys.\ Lett.\ B {\bf 679}, 504 (2009)
  [arXiv:0905.0751 [hep-th]];
  Y.~S.~Myung,
  %``Thermodynamics of black holes in the deformed Ho\v{r}ava-Lifshitz gravity,''
 Phys.\ Lett.\ B {\bf 678}, 127 (2009)
  [arXiv:0905.0957 [hep-th]].

%\bibitem{Cai:2009}
 % R.~-G.~Cai, B.~Hu and H.~-B.~Zhang,
  %``Dynamical Scalar Degree of Freedom in Ho\v{r}ava-Lifshitz Gravity,''
  %Phys.\ Rev.\ D {\bf 80}, 041501 (2009).
  %  [arXiv:0905.0255 [hep-th]].

\bibitem{Park:0910}
  M.~-I.~Park,
  %``Remarks on the Scalar Graviton Decoupling and Consistency of Ho\v{r}ava Gravity,''
  Class.\ Quant.\ Grav.\  {\bf 28}, 015004 (2011)
    [arXiv:0910.1917 [hep-th]].

\bibitem{Angh}
  M.~Angheben, %{\it et. al.},
  M.~Nadalini, L.~Vanzo and S.~Zerbini,
  %``Hawking radiation as tunneling for extremal and rotating black holes,''
  JHEP {\bf 0505}, 014 (2005)
    [hep-th/0503081].

\bibitem{Viss}
  M.~Visser,
 %``Hawking radiation without black hole entropy,''
  Phys.\ Rev.\ Lett.\  {\bf 80}, 3436 (1998)
   [gr-qc/9712016].

\bibitem{Deve}
  D.~O.~Devecioglu and O.~Sarioglu,
  %``On the thermodynamics of Lifshitz black holes,''
  Phys.\ Rev.\ D {\bf 83}, 124041 (2011)
   [arXiv:1103.1993 [hep-th]].

\end{thebibliography}
\end{document}